\documentclass[sigconf]{acmart}

\usepackage{algorithm}
\usepackage{algorithmic}
\usepackage{makecell}
\usepackage{tabularx}

\DeclareMathOperator{\softmax}{softmax}
\DeclareMathOperator{\leakyrelu}{LeakyReLU}

\AtBeginDocument{%
  \providecommand\BibTeX{{%
    \normalfont B\kern-0.5em{\scshape i\kern-0.25em b}\kern-0.8em\TeX}}}

\copyrightyear{2024}
\acmYear{2024}
\setcopyright{rightsretained}
\acmConference[KDD '24]{Proceedings of the 30th ACM SIGKDD Conference on Knowledge Discovery and Data Mining}{August 25--29, 2024}{Barcelona, Spain}
\acmBooktitle{Proceedings of the 30th ACM SIGKDD Conference on Knowledge Discovery and Data Mining (KDD '24), August 25--29, 2024, Barcelona, Spain}
\acmDOI{10.1145/3637528.3671562}
\acmISBN{979-8-4007-0490-1/24/08}

\makeatletter
\gdef\@copyrightpermission{
  \begin{minipage}{0.3\columnwidth}
   \href{https://creativecommons.org/licenses/by-nc/4.0/}{\includegraphics[width=0.90\textwidth]{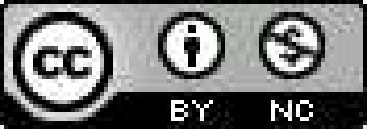}}
  \end{minipage}\hfill
  \begin{minipage}{0.7\columnwidth}
   \href{https://creativecommons.org/licenses/by-nc/4.0/}{This work is licensed under a Creative Commons Attribution-NonCommercial International 4.0 License.}
  \end{minipage}
  \vspace{5pt}
}
\makeatother

\begin{document}

\title{NudgeRank: Digital Algorithmic Nudging for Personalized Health}



\author{Jodi Chiam}
\authornote{Both authors contributed equally to this research.}
\affiliation{
  \institution{CueZen, Inc.}
  \city{Singapore}
  \country{Singapore}
}
\email{jodi.chiam@cuezen.com}

\author{Aloysius Lim}
\authornotemark[1]
\affiliation{
  \institution{CueZen, Inc.}
  \city{Singapore}
  \country{Singapore}
}
\email{aloysius.lim@cuezen.com}

\author{Ankur Teredesai}
\affiliation{
  \institution{CueZen, Inc.}
  \city{Seattle}
  \state{WA}
  \country{United States}
  \postcode{98402}
}
\affiliation{
  \institution{University of Washington}
  \city{Tacoma}
  \state{WA}
  \country{United States}
  \postcode{98402}
}
\email{ankurt@uw.edu}

\begin{abstract}
In this paper we describe NudgeRank™, an innovative digital algorithmic nudging system designed to foster positive health behaviors on a population-wide scale. Utilizing a novel combination of Graph Neural Networks augmented with an extensible Knowledge Graph, this Recommender System is operational in production, delivering personalized and context-aware nudges to over 1.1 million care recipients daily. This enterprise deployment marks one of the largest AI-driven health behavior change initiatives, accommodating diverse health conditions and wearable devices. Rigorous evaluation reveals statistically significant improvements in health outcomes, including a 6.17\% increase in daily steps and 7.61\% more exercise minutes. Moreover, user engagement and program enrollment surged, with a 13.1\% open rate compared to baseline systems' 4\%. Demonstrating scalability and reliability, NudgeRank™ operates efficiently on commodity compute resources while maintaining automation and observability standards essential for production systems.

\end{abstract}





\maketitle


\section{Introduction}\label{introduction}

Recommender Systems (RecSys) are pivotal and have been deployed in production across various industries, including e-commerce \cite{linden2003amazon}, streaming services \cite{covington2016deep}, music \cite{anderson2020algorithmic}, and social media \cite{ying2018graph}. They enable users to discover relevant items amidst overwhelming choices, enhancing engagement and satisfaction. This, in turn, drives growth for platforms and their user base.

In healthcare, RecSys play a crucial role, aiding both clinicians and patients. For example, they support clinical decision-making with drug recommendations considering patient history of allergies and drug interactions \cite{doulaverakis2014panacea, chen2011recommendation}, facilitate finding suitable healthcare providers \cite{zhu2021association}, and enable access to relevant health information \cite{sanchez2017healthrecsys}.

Despite these advancements, leveraging RecSys for \textit{digital algorithmic nudging} to promote healthy lifestyles across a large population remains underexplored. With the ubiquity of health devices and wearables, the data collected—ranging from physical activity (steps, active exercise minutes) to vital signs (heart rate, oxygen saturation, blood glucose)—offers a comprehensive view of an individual’s health behavior and states. Such data is used, with consent, to motivate health tracking and behavior modification through \textit{nudges}, a concept rooted in behavioral economics about choices presented to guide individuals towards desired behaviors \cite{Thaler_Sunstein_2009}. Several wearable device manufacturers use rule based systems to generate nudges that aim to guide the consumer to focus on certain activities. In general health and wellness settings, such nudges can encourage users to exercise more, get enough sleep, quit smoking and reduce alcohol intake, and even help develop healthy eating habits. Similarly in clinical healthcare settings nudges can help increase patient engagement for improving outcomes such as medication adherence, increasing screening rates for early detection of cancers, and encouraging clinical-trial patients to stay on protocol. In general, some level of nudging is recognized as useful for most people, whether they are healthy, or if they have a noncommunicable disease (NCD) like diabetes or cardiovascular disease, where behavior modifications are critical for preventing disease progression and complications. At a population level, more individuals adopting healthy behaviors can reduce the incidence and burden of NCDs, which are responsible for 74\% of deaths globally \cite{world2022noncommunicable, world2023noncommunicable}.

Achieving such sustained behavior change at a large population scale presents two notable challenges from socioeconomic and technical perspectives. 

First, changing human behavior is hard. People are creatures of habit. To be effective, nudges need to be \textit{personalized} to the recipient and \textit{contextual} to their circumstances. Nudges also have to be perceived to be coming from a human expert, and must be relevant and useful. In contrast to traditional outreach methods like billboards and mass emails, a RecSys can provide a much more private and intimate user experience by sending nudges on a regular basis and saying the appropriate thing at the right time, which can ultimately contribute to sustained behavior change.

Second, implementing a RecSys for personalized health nudging at population scale requires the system to address numerous technical challenges. It has to be highly compliant to care protocols, while being scalable and performant to generate recommendations for millions of users in real-time, or hourly, or daily. It needs to be easily configurable to meet specific business and compliance requirements of each local region and a variety of health goals. Such a system also needs to be reliable to operate in a production setting without any major downtimes and failures given the clinical importance of the interactions. It must be fully automated to gather performance and health metrics continuously; and capable of addressing issues like new user and new health goal cold start in a health setting where traditional methods may not apply directly.

In this paper we describe one such system we developed termed NudgeRank\texttrademark{}: a RecSys for digital algorithmic nudging for health. It is currently deployed in Singapore through a mobile health application to support the population's health and wellness goals.  NudgeRank\texttrademark{} has been running in production since July 2022, and scales to send digital personalized nudges to over 1.1 million active users of a mobile health application. Users sync their fitness trackers to the application and monitor their progress in various health and wellness programs, including physical activity like steps and moderate to vigorous physical activity (MVPA), and nudges are delivered via push notifications. We evaluated the effectiveness of algorithmic nudging in engaging users as well as efficacy of enabling changes in health behaviors, model performance, and overall system performance.

The main contributions of this work are:

\begin{itemize}
    \item  We describe a novel Graph Neural Network (GNN) based digital algorithmic nudging recommendation system for personalized health, with deployment details and results that can be applied to a wide range of health use cases and behaviors.
    \item We complemented the GNN with an extensible Knowledge Graph framework to capture rich information about care recipients' states and behaviors coming from lifestyle and health data sources like wearables and health records, to contextualize and personalize the nudges for individuals.
    \item We deployed the system to encourage awareness of improving physical activity levels in a large population (> 1 million users) in Singapore, and demonstrated its efficacy at health metrics such as improving step count and MVPA across the population.
    \item We demonstrated that the deployed system meets stringent compliance criteria for nudges, and is able to handle the scalability and reliability requirements of the overall system to operate in production at population scale.
\end{itemize}

\section{Related Work}\label{related-work}

The landscape of health behavior nudging encompasses a spectrum of strategies, from basic environmental cues to sophisticated digital interventions. For decades, policy makers have encouraged approaches that have utilized methods such as strategic placement of milk and vegetables to promote nutritional needs of consumers \cite{mistura2019examining}, used socially targeted messages on shopping trolleys to increase vegetable purchases \cite{huitink2020social}, and the alteration of trash bag colors to improve waste disposal practices \cite{abdel2023zero} for example. Despite their merits, these methods lack personalization and context-awareness necessary for broader applicability.

Advancements in consumer technology and Just-In-Time Adaptive Interventions (JITAIs) signify a shift towards leveraging mobile and sensor technologies for real-time data collection about users' changing internal and contextual states, enabling timely and relevant nudges \cite{nahum2018just}. These interventions span such a variety of health domains, including physical activity \cite{consolvo2008activity}, alcohol cessation \cite{gustafson2014smartphone}, smoking cessation \cite{riley2008internet}, and weight loss \cite{patrick2009text} that it is impossible to cite all the relevant studies here and we point the reader to but few of the ones we found interesting.  For example, Consolvo et al., used a personalized display to visualize the user's progress towards daily physical activity goals, but did not send any messages\cite{consolvo2008activity}. Gustafson et al., addressed this and applied decision rules to send text messages: to send an alert if the user's GPS location is in the proximity of a high-risk area for alcohol consumption~\cite{gustafson2014smartphone}. Riley et al., send messages as users progress through different stages of their care plans~\cite{riley2008internet}, and Patrick et al. incorporate a library of 1,500 decision rules to decide which message to send and when, based on user's state~\cite{patrick2009text}. While these interventions are context-sensitive, their reliance on a universal set of decision rules limits their personalization capacity.

Application of Machine Learning (ML) has increasingly enabled personalization of health interventions to adjust to individual preferences. Notably, the HeartSteps app employs Reinforcement Learning to tailor the timing of nudges \cite{liao2020personalized}, while other studies utilize hierarchical clustering and binary classification to refine daily goal recommendations~\cite{li2018adaptive} and employ LSTM models for exercise recommendations based on predicted heart rate profiles \cite{ni2019modeling}. These approaches were designed to offer study-specific personalization for a single goal and don't report results across simultaenous multiple health outcomes, an important criteria necessary to address poly-chronic and multiple comorbidities that exist in any population which our deployed system addresses in an extensible way.

We demonstrate that an AI driven extensible RecSys can offer a versatile foundation for developing personalized nudge systems applicable across various health scenarios. We were motivated by other preliminary successes in targeted studies, such as smoking cessation efforts achieving significantly higher abstinence rates with personalized messaging~\cite{hors2018tailoring, carrasco2020mobile}.

Based on the demonstrated commercial success of RecSys in significant operational contexts across various industries, we posited that a newly devised RecSys tailored for healthcare could sufficiently meet the requirements to establish a population-scale nudging mechanism. Our inquiry is predicated on the notion that RecSys, particularly those enhanced with knowledge graphs, possess the capability to adeptly expand personalized health nudging initiatives. Knowledge graphs offer the flexibility to assimilate a wide array of health-related information, encompassing data from fitness tracker activities to various smart sensors and clinical records, without necessitating extensive feature engineering efforts. This approach not only enriches the contextual basis for personalization but also leverages semantic relationships for more nuanced recommendations. Examples of knowledge-enhanced RecSys include RippleNet~\cite{wang2018ripplenet} and KGCN \cite{wang2019knowledge}, which utilize knowledge graphs for item recommendations, KGAT~\cite{wang2019kgat}, KNI~\cite{qu2019end}, which integrate user-item interactions for a comprehensive understanding of user preferences, and IntentGC~\cite{zhao2019intentgc} which captures knowledge about users.

To the best of our knowledge, this paper introduces the first GNN and Knowledge Graph augmented deployment of a RecSys for nudging for health that is capable of dynamically generating personalized nudges for millions of users. Our system design described in the next section demonstrates the practical application and effectiveness of this approach for real-world enterprise health setting.

\section{System Design}

NudgeRank\texttrademark{} by design is agnostic to specific health scenarios. While we present the system design and evaluation of nudging for population health promotion, it is also being used in other use cases where it is important to sustain desired health behaviors, like chronic disease management and pre- or post-surgical care.

\subsection{NudgeRank System}\label{nudgerank-system}

\begin{figure*}
    \centering
    \includegraphics[width=1\linewidth]{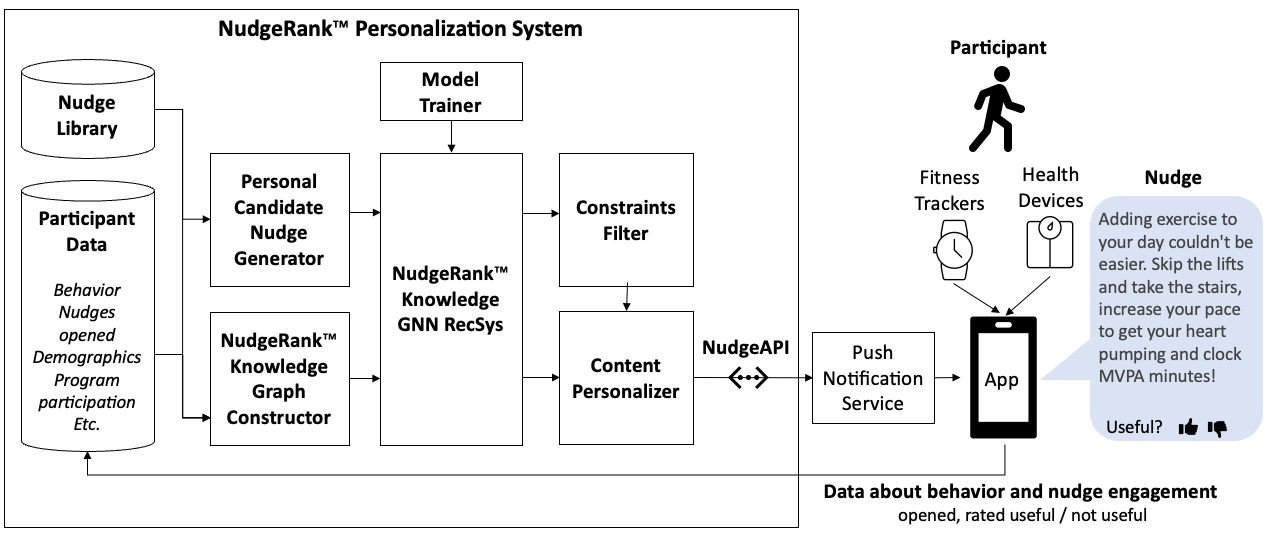}
    \caption{NudgeRank\texttrademark{} system that automatically generates personalized health nudges for millions of users daily, integrated with mobile app to deliver of nudges via push notifications and collect contextual and nudge engagement data.}
    \Description{The diagram shows a schematic and main components of the NudgeRank\texttrademark{} system, including Nudge Library, Participant Data, Personal Candidate Nudge Generator, NudgeRank\texttrademark{} Knowledge Graph Constructor, NudgeRank GNN RecSys, Model Trainer, Constraints Filter, Content Personalizer, and NudgeAPI.}
    \label{fig:nudgerank-system}
\end{figure*}

The NudgeRank\texttrademark{} system is depicted in Figure \ref{fig:nudgerank-system}, highlighting its primary components.

\subsubsection{Data Collection: Nudge Library and Participant Data}\label{nudge-library-participant-data}

The system commences by aggregating essential data. The \textit{Nudge Library} enables care program manager to author nudge templates that can be personalized for end users. \textit{Participant data}, encompassing demographics, health behaviors, and more, are sourced via API from mobile apps or other databases, including electronic health records. This dataset is enriched by user-item interactions, capturing engagement with the nudges, including opening or rating them (e.g. thumbs up or down).

\subsubsection{Knowledge Graph Neural Network Recommender System}\label{kg-recsys}

Data preparation for the RecSys involves the \textit{Personal Candidate Nudge Generator} generating a tailored list of candidate nudges for each user based on predefined targeting rules. For example, some nudges may only be targeted at users over 60 years old who have walked less than 5,000 steps in the previous week.

\begin{figure*}
    \centering
    \includegraphics[width=1\linewidth]{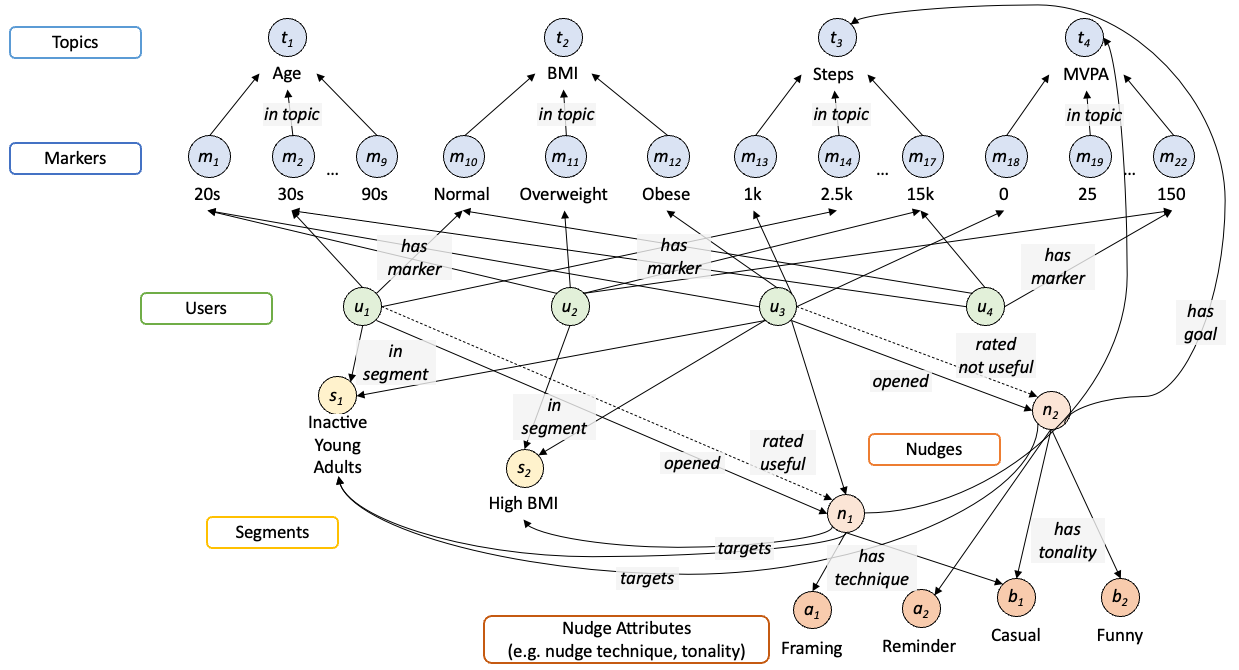}
    \caption{NudgeRank\texttrademark{} Knowledge Graph.}
    \Description{The figure shows examples of entities and relations captured in the NudgeRank\texttrademark{} Knowledge Graph.}
    \label{fig:nudgerank-kg}
\end{figure*}

\begin{figure*}
    \centering
    \includegraphics[width=1\linewidth]{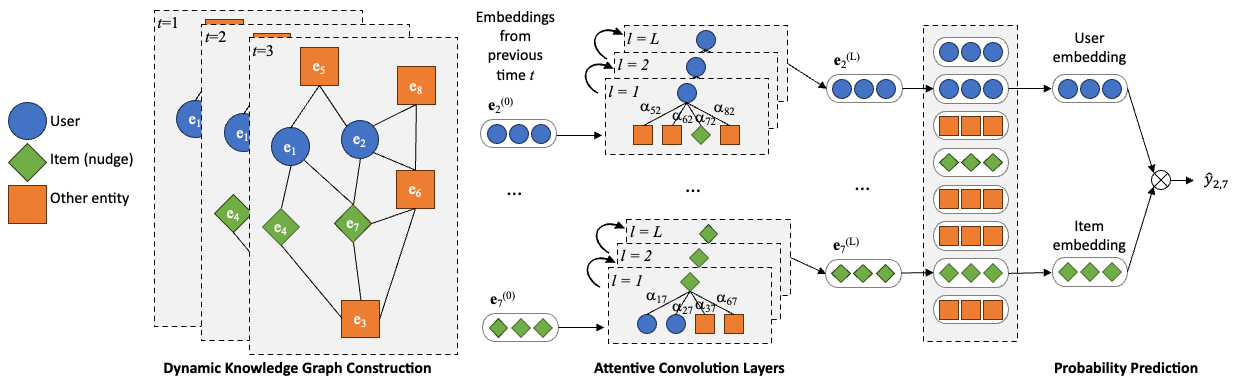}
    \caption{NudgeRank\texttrademark{} Knowledge Graph Neural Network.}
    \Description{The architecture of the NudgeRank RecSys, including dynamic graph update, attentive convolution layers, and prediction layer.}
    \label{fig:nudgerank-gnn}
\end{figure*}

The \textit{NudgeRank\texttrademark{} Knowledge Graph Constructor} forms a heterogeneous knowledge graph, capturing the complex interplay between users, nudges, and their attributes. This graph dynamically updates in response to new user data, ensuring recommendations reflect the current contextual state of each user. Users and nudges are represented as nodes in the graph, and directed edges capture the history of user's past interactions with the nudges. In the example in Figure \ref{fig:nudgerank-kg}, user \(u_1\) opened nudge \(n_1\) and rated it "useful". Users' attributes and behaviors are represented by binary \textit{markers}, which are classified into \textit{topics}. For example, user \(u_1\) has the markers \textit{"age: 30s"}, \textit{"BMI: normal"} and \textit{"steps: 2.5k"}. While the figure only illustrates a few markers, the actual graph contains over 130 markers, and can be easily extended with more data sources and adapted to more use cases, e.g. HbA1C or blood glucose markers for diabetes. The graph also captures knowledge about the nudges. For example, nudge \(n_2\) targets segment \(s_1\) \textit{"Inactive Young Adults"}, and was written to encourage them to walk more steps.

We extend prior efforts (\cite{wang2018ripplenet, wang2019knowledge, wang2019kgat, qu2019end, zhao2019intentgc}) in two important ways for algorithmic digital nudging in health:

First, besides capturing knowledge about items, the knowledge graph also captures knowledge about users, such as their demographics, health behaviors and segments. Furthermore, it captures shared knowledge between users and items, in addition to their interaction histories. For example, in Figure \ref{fig:nudgerank-kg}, users \(u_1\) and \(u_3\) are in segment \(s_1\) since they are both in their 20's and 30's, and have not been active in either steps or MVPA. Nudge \(n_2\) targets segment \(s_1\) and has a "steps" goal, and would therefore be a candidate nudge to encourage \(u_1\) and \(u_3\) to walk more steps. By capturing all the knowledge about users and nudges and the relationships between them, the knowledge graph provides a rich representation of the world for the RecSys to finely tailor nudges according to each user's preferences and contextual states.

Second, the graph is \textit{dynamic} and updated frequently as new information about users are collected, to ensure that recommendations are always made in the context of the users' current states and behaviors. For example, if a user had the marker \textit{"steps: 2.5k"} one week ago, but the marker changed to \textit{"steps: 10k"} this week because the user walked more, the user's graph connections are updated to reflect this new information, and new nudges are generated with the updated graph.

Both the list of candidate nudges per user and the knowledge graph are fed to the RecSys, which ranks the candidate nudges for each user, based on higher-order relationships between users and items captured in the knowledge graph.

Let $\mathcal{U}$ be the set of users, $\mathcal{I}$ be the set of items (nudges), $\mathcal{E^\prime}$ be the set of other entities, and $\mathcal{R}$ be the set of relation types. Then, the dynamic NudgeRank\texttrademark{} Knowledge Graph at time $t$ is defined by $\mathcal{G}_t = \{(a,r,b)|a,b \in \mathcal{E}, r \in \mathcal{R}\}$, where $\mathcal{E} = \mathcal{U}\cup\mathcal{I}\cup\mathcal{E}^\prime$ and a triplet $(a,r,b)$ represents a relationship $r$ from head node $a$ to tail node $b$ that exists at time $t$. For example, the triplet $(u_1, in\_segment, s_1)$ is an edge in the graph indicating that user $u_1$ is in segment $s_1$. Each node in $\mathcal{E}$ is represented by a $d$-dimensional embedding vector $\mathbf{e} \in \mathbb{R}^d$, and each relation in $\mathcal{R}$ is represented by a $k$-dimensional embedding vector $\mathbf{r} \in \mathbb{R}^k$.

For initial training, all node and relation embeddings are initialized using Xavier initialization \cite{glorot2010understanding}. When the model is subsequently fine-tuned on new data, the node and relation embeddings of graph $\mathcal{G}_{t+1}$ are first initialized using the previously learned embeddings from graph $\mathcal{G}_t$, except for any new nodes or relations where Xavier initialization was applied.

The NudgeRank\texttrademark{} model (Figure \ref{fig:nudgerank-gnn}) uses attentive graph convolution layers to aggregate node embeddings from neighboring nodes, so that the GNN can learn which neighbors are most important for passing information to personalize the nudges. The number of layers $L$ and the dimensions of each layer are hyperparameters that can be tuned. The embeddings for head node $a$ in layer $l$ are updated by the attention-weighted linear combination of its first order neighbors $\mathcal{N}_a$ in layer $l - 1$, aggregated with its own embeddings via a differentiable function $f$, e.g. addition or concatenation.

\begin{equation}\label{eq:propagation}
    \mathbf{e}_{a}^{(l)} = f\left(\mathbf{e}_{a}^{(l-1)}, \sum_{b\in \mathcal{N}_a} \alpha_{ab} \mathbf{e}_{b}^{(l-1)} \right)
\end{equation}

We adopted Knowledge-Aware Attention \cite{wang2019kgat}, where the relation-specific attention weights $\alpha_{ab}^r$ are defined in Equation \ref{eq:kgat-attention}, with matrix $\mathbf{W}_r$ projecting the node embeddings $\mathbf{e}_a$ and $\mathbf{e}_b$ to the space of relation $r$. Thus, the attention weight for the relation $(a,r,b)$ is proportional to the similarity between $\vec{a}+\vec{r}$ and $\vec{b}$ in the space of relation $r$.

\begin{equation}\label{eq:kgat-attention}
    \alpha_{ab}^r = \softmax_{b}((\mathbf{W}_r \mathbf{e}_b)^\top \tanh(\mathbf{W}_r \mathbf{e}_a + \mathbf{e}_r))
\end{equation}

The NudgeRank\texttrademark{} framework is extensible to allow other attention mechanisms to be used, for example, the Graph Attention Layer \cite{velivckovic2017graph} (Equation \ref{eq:gat}), which includes self-attention and supports multiple attention heads. We left alternate attention mechanisms for future exploration.

\begin{equation}\label{eq:gat}
    \alpha_{ab} = \softmax_b(\leakyrelu(\mathbf{a}[\mathbf{We}_a \Vert \mathbf{We}_b]))
\end{equation}

After node embeddings have propagated through the convolutional layers, the predicted rating for user $u$ and item $i$ is computed as the dot product of the embedding vectors:

\begin{equation}
    \hat{y}_{u,i} = \mathbf{e}_u^\top \mathbf{e}_i
\end{equation}

\subsubsection{Constraints Filter}\label{constraints-filter}

After the RecSys has ranked the candidate nudges for each user, the \textit{Constraints Filter} applies customizable criteria to refine the nudge selection, aiming to optimize engagement while mitigating fatigue. For example, a nudge budget (maximum number of nudges per day) is applied to minimize nudge fatigue; nudges that were sent recently (e.g. last 7 days) are removed to increase the diversity of nudges in consecutive days; nudges that a user had rated "not useful" are removed to reduce annoyance. All of these nudges are fully configurable according to the business needs of the health program.

\subsubsection{Content Personalizer}\label{content-personalizer}

The \textit{Content Personalizer} processes the final list of selected nudges for each user, and renders any personalized fields in the nudge template. For example, the following nudge template \textit{"Great job walking \{\{avg\_daily\_steps\}\} daily steps last week! You are getting pretty close to your goal of 10,000 steps per day!"} is rendered with the user's actual data (e.g. \textit{"8,356 daily  steps"}) to personalize the nudge for that user.

\subsubsection{NudgeAPI}\label{nudge-api}

Finally, the nudges are exposed via \textit{NudgeAPI} to external delivery channels such as mobile apps or email delivery systems. Feedback data in the form of nudge engagement (sent, opened, ratings) are collected, along with health behavior and vitals from fitness trackers and other smart devices. As new data is ingested, they create a continuous feedback loop with updated contextual information for the next set of personalized nudges to be generated.

\subsection{Design Considerations}\label{design-considerations}

Deploying a RecSys at scale requires careful design to solve operational issues that can occur in a production system. This section describes the critical design considerations for NudgeRank\texttrademark{}.

\subsubsection{Personalization and Contextualization}\label{personalization-contextualization}

The main job of a RecSys for health nudges is to deliver personalized nudges that are relevant and contextual to the user's health state, goals and behaviors, so that the user remains engaged with the health program and adheres to the desired positive health behaviors. NudgeRank\texttrademark{} achieves a high level of personalization by considering \textit{what} to say, \textit{who} to send a nudge to, and \textit{when} to send it.

\textit{What}: the Nudge Library allows a wide variety of nudges to be written, using different nudge techniques, choice of words, or tonalities. Since different users may find different nudges to be interesting, NudgeRank\texttrademark{} can learn the preferences of individual users and recommend other similar nudges. For example, some users may have a competitive personality and respond better to nudges that challenge them with social comparisons to peer groups, while others respond better to gentle reminders.  Furthermore, nudges can be written as \textit{templates} that allow the content to be personalized by injecting the user's own contextual data into the message, as described in Section \ref{content-personalizer}.

\textit{Who}: The program manager can write nudges for different audiences by specifying the target \textit{segments} for each nudge. For example, different nudges can be written for young adults who are sedentary vs. active. Segments can be defined by static user attributes like demographics, and dynamic attributes like health behaviors. A user's segment memberships are dynamic and automatically updated, as their health states and behaviors change daily. As a result, the list of candidate nudges that the user may receive is also contextual and up-to-date. For example, suppose the user belongs to the "Inactive Young Adult" segment on day 1, and gets a nudge reminding the user to exercise. On day 2, the user goes to the gym and moves to another segment called "Active Young Adult"; the user receives a different nudge saying, "Good job, keep it up!" Segments can be defined as broadly or precisely as required; for example, nudges can be written for all "Young Adults", or only "Prediabetic Overweight Moderately Active Young Males". As users perform different health behaviors or their health states change, dynamic segments ensure that the right nudges are selected for them.

\textit{When}: Underpinning the ability to send the right nudge at the right time is the \textit{dynamic} knowledge graph that provides the latest state of the world for the RecSys model to generate nudges with. As new data is collected about users and how they have interacted with nudges, the knowledge graph's nodes and edges are updated with the latest data. In this way, recommendations are \textit{timely} and relevant to each user's current context and behaviors.

\subsubsection{Cold Start}\label{cold-start}

In NudgeRank\texttrademark{}, the cold start problem can happen when (a) new nudges are added to the Nudge Library, or (b) new users enroll for the health program and they are added to the graph. We solve cold start in two main ways.

First, new users and nudges are added to the knowledge graph, with any information that is known about them. Users have demographic or other data that is collected about them during enrollment. Nudges minimally have a target health goal, target segments, and other attributes. Using these basic information, graph connections can already be established for new users and nudges. Node embeddings representing new users and nudges are initialized using the Xavier initializer \cite{glorot2010understanding}, which effectively randomizes the new users and nudges without any learned preferences. By adding new users to the graph, the RecSys is able to recommend nudges that are relevant to the users, even in the absence of any behavior data or nudge interactions; for example, nudges that were opened by other users in the same demographic groups. As new users record their health behaviors and interact with nudges, the recommendations become more and more personalized to their individual contexts and preferences. Likewise, new nudges with no interaction history can already be recommended to users in the relevant target segments, and based on other graph connections.

Second, we introduced a mechanism to ensure \textit{diversity} of nudges. A configurable parameter $p_{diversity} \in [0.0, 1.0]$ controls the proportion of nudges that are randomly sampled without replacement from each user's list of candidate nudges. For every nudge that is output by the Constraints Filter (Section \ref{constraints-filter}), there is a probability of $p_{diversity}$ that the nudge is replaced by a uniformly sampled nudge from the user's candidate nudge list. In doing so, some of the nudges actually sent to the user may have been below the top $k$ ranked nudges. This allows a greater diversity of nudges to be sent, including new nudges which are typically not ranked high. Increasing diversity also allows for more user-nudge interaction data to be collected, allowing the RecSys to learn user preferences more broadly.

\subsubsection{Business Rules and Constraints}\label{business-rules-constraints}

To ensure and maintain a positive user experience, critical business rules can be configured in the system, according to the specific needs of the health program. These are enforced by the Constraints Filter (Section \ref{constraints-filter}).

A \textit{Negative Rating Filter} removes all nudges that the user has given a negative rating for (e.g. thumbs down) in the last $d_{neg\_filter} \geq 0$ days. This prevents such nudges from being recommended again, to maintain a positive user experience. The nudge can be sent again after $d_{neg\_filter}$ days, in case the user's preferences or context have changed.

Because nudges are delivered via push notifications, a \textit{Nudge Budget Filter} allows the number of daily nudges to be limited to a configurable parameter $k_{daily}$, to reduce nudge fatigue and annoyance to users. If a user has $k_{recommended}$ recommended nudges after all other constraints have been applied, then the actual number of nudges sent is $k_{final} = \min(k_{daily}, k_{recommended})$.

\subsubsection{Performance and Resilience}\label{performance-resilience}

Generating nudges for a large population in production requires the NudgeRank\texttrademark{} system to be engineered with performance and resilience in mind. NudgeRank\texttrademark{} runs on Kubernetes, with critical pipelines running in parallel and distributed across the cluster. This allows the system to scale horizontally on commodity infrastructure, including virtual machines and storage from any cloud provider. Specifically, the batch scoring pipelines run in parallel by processing and generating different subsets of the population. As many instances of the pipeline as needed can be instantiated to handle the load of a large and growing population of users. Based on the number of Kubernetes nodes available and the desired overall execution time, the parameter $b > 0$ determines how many parallel batches are processed. The $n$ users are divided into $b$ equal batches, then the NudgeRank\texttrademark{} pipeline is run for each batch of users. The generated nudges for all users are then concatenated at the final step before publishing the nudges in NudgeAPI. This is illustrated in Algorithm \ref{alg:parallel-nudgerank}.

A distributed system architecture running in the cloud introduces sources of potential failures in the system. For example, intermittent network issues when reading or writing data, or issues in the Kubernetes cluster may cause various parts of the pipeline to fail. To improve the resilience of the system, an automated retry mechanism was built into the parallel algorithm, to automatically recover from failures and increase the fault-tolerance of the system. If any failures occur within a batch resulting in an exception being raised, the processing for that batch is repeated. Because batches contain non-overlapping subsets of users, there is no conflict with other batches being processed. To minimize the time to recover from failures, only failed batches need to be re-executed, and the results of other successful batches are retained for the final output.

\begin{algorithm}
\caption{Parallel NudgeRank\texttrademark{}}\label{alg:parallel-nudgerank}
\begin{algorithmic}
\REQUIRE $n > 0$ and $b \geq 1$
\STATE $nudges \gets \{\}$
\STATE $n_{batch} \gets \lceil \frac{n}{b} \rceil$
\STATE $\mathcal{G} \gets$ KGConstructor($nudge\_library$, $participant\_data$)
\FORALL{$q, 1 \leq q \leq b$}
    \STATE /* Run all batches in parallel. If any batch fails, repeat it. */
    \STATE $batch\_users = \{u_i | (q-1) \cdot n_{batch} < i \leq q \cdot n_{batch}\}$
    \STATE $candidate\_nudges \gets \{\}$
    \FORALL{$u \in batch\_users$}
        \STATE $candidates \gets$ CandidateNudgeGenerator($u$)
        \STATE $candidate\_nudges \gets candidate\_nudges \cup \{(u, i) \forall i \in candidates\}$
    \ENDFOR
    \STATE $ranked\_nudges \gets$ NudgeRankRecSys($\mathcal{G}$, $candidate\_nudges$)
    \STATE $filtered\_nudges \gets$ ConstraintsFilter($ranked\_nudges$)
    \STATE $personal\_nudges \gets$ ContentPersonalizer($filtered\_nudges$)
    \STATE $nudges \gets nudges \cup personal\_nudges$
\ENDFOR
\RETURN $nudges$
\end{algorithmic}
\end{algorithm}

\subsubsection{Operations and Automation}\label{operations-automation}

To streamline production operations, the key pipelines in the system are fully automated. DevOps Engineers and Data Scientists can monitor the system and be alerted of any issues, in order to investigate and fix them quickly. The Model Trainer automatically fine-tunes the model daily as the graph gets updated with new data about user states and nudge interactions, to ensure that model learns about user preferences in a timely manner. This was especially critical in the cold start period when the system was first launched and users interacted with nudges for the first time. Batch scoring as described in the preceding section is also automated on a schedule, as data is ingested from the source systems. Finally, all pipelines capture system telemetry including graph attributes (e.g. number of nodes and edges, number of users, number of nudges), pipeline performance (e.g. execution time) and model performance during training and fine-tuning (e.g. recommendation metrics).

\subsubsection{Security and Privacy}\label{security}

Privacy and security of information is paramount when dealing with health data, especially in cases where the data comes from medical records. To ensure that the system complies with global data protection standards for health data, NudgeRank\texttrademark{} is designed to work without any Personal Identifiable Information (PII) such as names or addresses. All user IDs in the system are pseudonymous masked IDs, and the system does not require any PII to perform its functions.

Furthermore, the system is designed to be deployed in "air gapped" environments with private networks and no public Internet access (including private networks in the cloud), to ensure that there is no risk of data egress. Data belonging to a healthcare provider or health program never leaves their environment; rather, NudgeRank\texttrademark{} is deployed in their environment with automated CI/CD for deployments, and remote monitoring of aggregated and anonymous system telemetry for daily operations.

\section{Results}\label{results}

We evaluated NudgeRank\texttrademark{} in a large-scale deployment in Singapore. The system was launched in production in July 2022 as part of several beta tests involving selected users. In July 2023, the system was launched for all users, and has since been generating personalized nudges for over 1.1 million users daily. All users had previously given consent for their fitness data to be collected via the mobile health application, and for their data to be used to send personalized messages, nudges or notifications, and for evaluation of results.

The system was deployed in a secure Azure cloud environment, with all data required for model updates and generating recommendations remaining within the environment. A 10-node Kubernetes cluster was used, with each node having 16 vCPUs and 128 GB RAM. Data was stored in Azure Data Lake Gen2, a cloud storage service supporting blobs in a hierarchical namespace.

With input from business requirements, the system was configured with the following parameters (described above): $b=8$, $k_{daily}=1$, $p_{diversity}=0.3$, and $d_{neg\_filter}=7$.

\subsection{Model Performance}\label{model-performance}

In the first limited beta phase, nudges were sent to about 100,000 randomly selected users for a period of 4 months, and the interaction data were collected to tune the hyperparameters of the model. The following hyperparameters were evaluated using a grid search: the dimension size of entity and relation embeddings, and the number of graph convolution layers and their respective embedding sizes of the hidden layers. For each combination of hyperparameters, the model was trained on the knowledge graph with 25\% of user-nudge interactions randomly hidden, and tested on the knowledge graph including those 25\% unseen interactions.

We theorized that behavior change via nudging starts with simply interacting with the nudges by opening or rating them. The more nudges are opened, the more likely it is that users have read them, and therefore make a choice about their behaviors. Therefore, offline evaluation focused on the model's ability to recommend top 3 daily nudges that users were most likely to interact with. $Precision@3$ was chosen as the primary evaluation metric, since it represents the proportion of the 3 recommended nudges that lead to user interaction.

Table \ref{tab:hyperparameter-tuning} shows the list of parameters tested, with the parameter combination that produced the best offline evaluation score of $Precision@3 = 0.0652$. This parameter combination was set up in the system for subsequent nudge recommendations. Table \ref{tab:offline-metrics} compares this with the best-performing parameter combinations for 1-, 2- and 3-layer models. The 2-layer model achieved the highest $Precision@3$; notably, this is 0.0201 (45\%) higher than the 1-layer model that serves as a baseline for comparison.

\begin{figure*}
    \centering
    \includegraphics[width=1\linewidth]{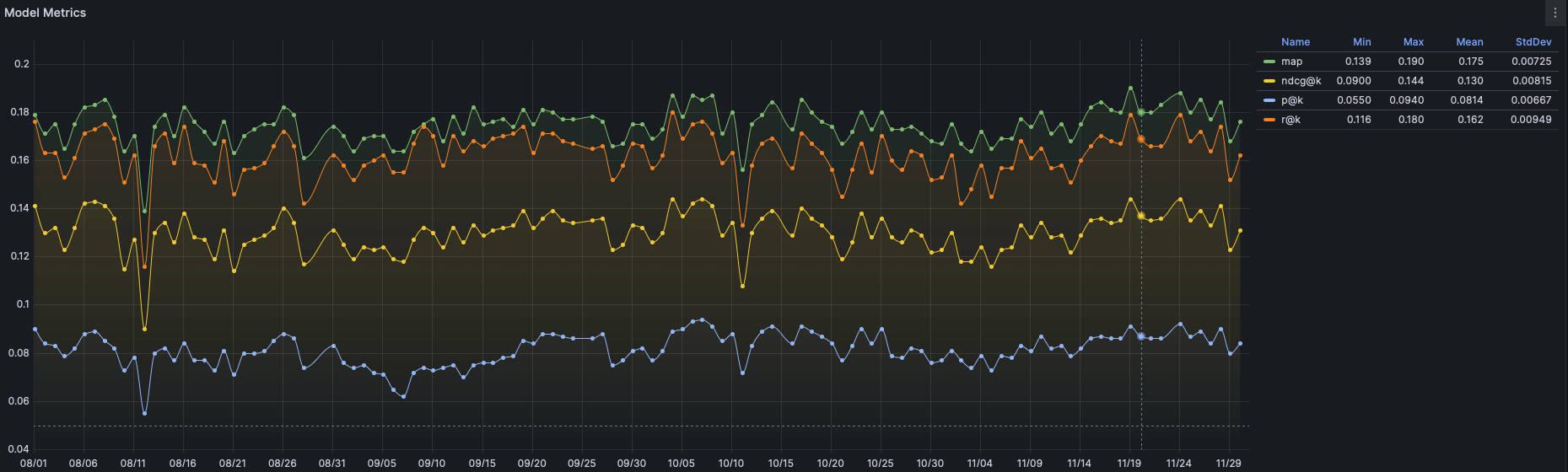}
    \caption{Evaluation metrics from daily automated model updates, showing the minimum, maximum, mean and standard deviation of Mean Average Precision (map), Normalized Discounted Cumulative Gain @ 3 (ndcg@k), Precision @ 3 (p@k) and Recall @ 3 (r@k).}
    \Description{A screenshot from model metrics monitoring dashboard showing the daily changes in model metrics.}
    \label{fig:daily-training-metrics}
\end{figure*}

After nudges were rolled out at full scale, we added more metrics ($Mean Average Precision$, $NDCG@3$ and $Recall@3$), and continued to monitor the offline metrics during daily incremental model updates. The screenshot in Figure \ref{fig:daily-training-metrics} was captured from the system's monitoring dashboards, showing that model metrics experienced daily fluctuations within fairly tight ranges. The standard deviations of the metrics ranged from 0.00684 for $Precision @ 3$ to 0.00996 for $Recall @ 3$ over the 4 month period from August 2023 to November 2023, indicating that model performance was fairly stable over time.

\begin{table*}
  \caption{Hyperparameters tested, and final best hyperparameters.}
  \label{tab:hyperparameter-tuning}
  \begin{tabular}{ccc}
    \toprule
    Hyperparameter& Search space& Selected Best Value\\
    \midrule
    Dimension of entity embeddings& 16, 32, 64& 32\\
    Dimension of relation embeddings& 16, 32, 64& 32\\
    Convolution layers \& dimension of hidden layers& \makecell{1 layer: [16], [32], [64]\\ 2 layers: [32, 16], [64, 32]\\ 3 layers: [64, 32, 16], [32, 16, 8]}& 2 layers: [32, 16]\\
  \bottomrule
\end{tabular}
\end{table*}

\begin{table*}
  \caption{Optimal parameters and offline evaluation results of the best 1-, 2-, and 3-layer models.}
  \label{tab:offline-metrics}
  \begin{tabular}{ccccc}
    \toprule
    No. of Layers& Dimension of entity embeddings& Dimension of relation embeddings& Dimension of hidden layers& $Precision@3$\\
    \midrule
    1& 64& 64& [64]& 0.0451\\
    2& 32& 32& [32, 16]& \textbf{0.0652}\\
    3& 64& 64& [64, 32, 16]& 0.0401\\
  \bottomrule
\end{tabular}
\end{table*}

\subsection{Impact of Nudging on Physical Activity}\label{impact-physical-activity}

Arguably the most important outcome to evaluate is whether NudgeRank\texttrademark{} is effective at improving health behaviors via nudging. We evaluated this over a 12-week period, where daily personalized nudges were sent to $n = 84,764$ randomly selected users of the mobile health application, to encourage them to do more physical activity including steps and MVPA. A matched control group of $n = 84,903$ users with the same demographics (age and sex) and baseline behaviors (mean steps and MVPA before nudging) were also selected to compare the effect of nudging. Users in the control group did not get any personalized nudges.

Over the 12-week period, users who received daily personalized nudges performed 6.17\% more average daily steps and 7.61\% more average weekly minutes of MVPA than users who did not receive any nudges. These results were statistically significant at $\alpha = 0.05$ on one-tailed independent samples t-tests comparing both groups (Table \ref{tab:nudge-impact}). When the behaviors of the nudge and no nudge groups were measured on a weekly basis for each of the 12 weeks, they remained significantly different for all 12 weeks for steps, and 10 out of 12 weeks for MVPA, suggesting that behavior differences were sustained and persistent.

Participants who opened more nudges were also found to have been more physically active. For example, participants who opened 0, 1, 2 and $\geq$ 3 nudges during the 12 weeks walked a mean of 2,491, 3,213, 3,588 and 4,445 daily steps, respectively. This suggests a positive relationship between nudge engagement and behavior.

\begin{table}
  \caption{Impact of nudging on physical activity. $p$-values were computed on one-tailed independent samples t-tests.}
  \label{tab:nudge-impact}
  \begin{tabular}{cccc}
    \toprule
    Group& \makecell{No Nudges\\(Control)}& \makecell{Daily Nudges\\(Treatment)}& $p$\\
    \midrule
    Daily Steps& 2,908.6& 3,088.1 (+6.17\%)& $3.09 \times 10^{-4}$\\
    Weekly MVPA& 47.0& 50.6 (+7.61\%)&$1.16 \times 10^{-2}$\\
  \bottomrule
\end{tabular}
\end{table}

Furthermore, nudges were generally well received. A total of 1.12 million nudges sent during this period, comprising 660,000 nudges to encourage steps and 460,000 nudges for MVPA. 13.1\% of all nudges were opened, out of which, 6 times more nudges were rated useful (11.7\%) than not useful (1.9\%) by users.

Appendix reports more representative results from this study on the impact of nudges on physical activity; for full coverage, see \cite{chiam2024copilot}.

\subsection{System Scalability and Performance}\label{system-scalability-performance}

On a daily basis, the production system generates nudges for over 1 million users. The typical size of the knowledge graph (which has small daily fluctuations) is reported in Table \ref{tab:kg-attributes}. In order to ensure that nudges are generated in time to be sent each day, a performance test was conducted to evaluate how total execution time increases with the number of candidate nudges that need to be ranked. Synthetic data was created to generate a larger number of candidate user-nudge pairs than was available in the production system, ranging from 300 thousand to 19 billion candidate pairs. For each set of candidate nudges, the system was executed with 8 parallel batches, with each batch process running on a Kubernetes node with 16 vCPUs and 128GB RAM. The total end-to-end execution time of Parallel NudgeRank\texttrademark{} (Algorithm \ref{alg:parallel-nudgerank}) was recorded.

\begin{figure}
    \centering
    \includegraphics[width=1\linewidth]{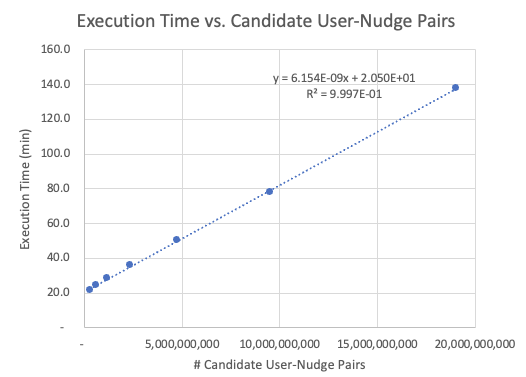}
    \caption{Results of execution time versus volume of candidate user-nudge pairs to be scored.}
    \Description{The results show that as the number of candidate pairs increases, the execution time increases linearly.}
    \label{fig:performance-scale}
\end{figure}

The results in Figure \ref{fig:performance-scale} show a strong correlation ($R^2 = 0.9997$) between the number of candidate user-nudge pairs and the total execution time of the system. Scoring 300 thousand candidate pairs on the tested infrastructure took 22 minutes. This grew linearly to 138 minutes to score 19 billion candidate pairs. The results indicate that the system is able to scale linearly to process increased demand of future workloads. Furthermore, the parallel design of the system makes it trivial to reduce the overall execution time, by adding more compute resources to the cluster.

Daily model updates were also computationally efficient in the production system, with the Model Trainer taking between 90 and 150 minutes to update the model with the latest daily knowledge graph. The variation in time was mainly due to the different number of epochs required by the model to converge, as data inputs changed day to day. The short time required for model update allows it to be scheduled daily, during system off-peak hours.

At the time of writing, the system has been running daily for over 18 months, indicating the high level of reliability achieved in its design.

\begin{table}
  \caption{Typical graph attributes in daily automatic model training.}
  \label{tab:kg-attributes}
  \begin{tabular}{cc}
    \toprule
    Attribute& Value\\
    \midrule
    Number of nodes& 3.1 million\\
    Number of edges& 5.7 million\\
    Density& $5.93 \times 10^{-7}$\\
  \bottomrule
\end{tabular}
\end{table}

\section{Discussion}\label{discussion}

We presented the design and implementation results of a production-scale RecSys for digital algorithmic nudging, and demonstrated the ability of the NudgeRank\texttrademark{} system to not only improve health behaviors, but also to do it at scale for a large population, and with a high level of performance and reliability. Although the system was first evaluated on physical activity (steps and MVPA), it is designed to be easily deployed for any other use cases where it is important to influence users' health behaviors, for example, diabetes management, nutrition, sleep, smoking cessation, pre- / post-operative care, and many others.

The system was launched to users in a progressive rollout, starting with fewer users and gradually increasing to the full user base over a period of 12 months. This allowed the system to be monitored and improved in iterations to address operational issues. For example, Parallel NudgeRank\texttrademark{} (Algorithm \ref{alg:parallel-nudgerank}) was developed in response to the increased load, as nudges were enabled for more users. The monitoring dashboards that recorded model performance and nudge engagement provided useful real-time insights into the system's daily operations for observability and troubleshooting. 

A number of future improvements are under development. Reinforcement Learning can be used to provide goal-directed recommendations, where nudges are based not only on user preferences, but also consider the "next best action" that a user should take towards stated health goals. The Knowledge Graph can be updated with more node and edge types, providing richer context for personalized recommendations. Large Language Models can be used to enrich the graph with semantic information, for example about nudge content, as well as add external knowledge about relevant health topics.

This work is just the first step in improving population-level health behaviors through personalization. As the future iterations of the system bring new capabilities and the system is deployed and evaluated on other health use cases and different populations, there is tremendous potential to effect sustained behavior change at scale, leading to meaningful improvements in population health outcomes.


\bibliographystyle{ACM-Reference-Format}
\balance
\bibliography{main}

\clearpage
\nobalance

\appendix

\section{Appendix: Population Health and Nudge Efficacy Results}

This section provides representative results of the study on impact of nudging on physical activity, namely steps and Moderate to Vigorous Physical Activity (MVPA).

For complete coverage and details, please refer to \cite{chiam2024copilot}.

\subsection{Background}

The mobile health application used in this study serves over 700,000 active participants in a physical activity program in Singapore, and over 1.1 million participants in a nutrition program that encourages healthy food purchases.

Traditionally, mass marketing efforts such as roadshows, broadcasts and social media were the main channels of engagement to encourage active participation in these programs. As part of a beta launch of NudgeRank\texttrademark{} integrated with the mobile health application, a study was conducted to evaluate the impact of digital algorithmic nudging on physical activity behavior such as steps and MVPA.

\subsection{Study Design}

\begin{table*}
    \centering
    \begin{tabularx}{\textwidth}{|X|X|X|X|X|} \hline 
         Group&  \multicolumn{2}{|>{\hsize=2\hsize}X|}{Group 1: Participants in physical activity program only}&  \multicolumn{2}{|>{\hsize=2\hsize}X|}{Group 2: Participants in both physical activity and nutrition programs} \\ \hline 
 Group n& \multicolumn{2}{|>{\hsize=2\hsize}X|}{14,901}& \multicolumn{2}{|>{\hsize=2\hsize}X|}{154,766} \\ \hline 
         Treatment&  No Nudges&  Nudges&  No Nudges&  Nudges\\ \hline 
         n&  7,465&  7,436&  77,438&  77,328\\ \hline 
         Mean age&  47.2&  47.6&  46.0&  46.0\\ \hline 
         \% Female&  48.0&  47.7&  62.9&  63.1\\ \hline 
         \% Male&  52.0&  52.3&  37.1&  36.9\\ \hline 
         Mean Daily Steps (start of experiment)&  3,138&  3,153&  4,154&  4,159\\ \hline 
         Mean Weekly MVPA (start of experiment)&  39.5&  40.0&  85.8&  86.8\\ \hline 
         \% iOS&  44.6&  44.5&  43.1&  43.2\\ \hline
         \% Android& 55.4& 55.5& 56.9& 56.8\\ \hline
         \% Apple Watch& 17.2& 16.1& 16.3& 16.5\\ \hline
         \% Fitbit& 6.2& 6.3& 9.8& 9.9\\ \hline
         \% Garmin& 8.0& 7.2& 6.3& 6.1\\ \hline
         \% Samsung Watch& 7.2& 6.9& 9.4& 9.3\\ \hline
         \% Program-issued Fitness Tracker& 61.0& 63.2& 56.8& 56.7\\ \hline
         \% Other Fitness Trackers& 0.4& 0.3& 1.4& 1.5\\ \hline
    \end{tabularx}
    \caption{Participants in control and treatment groups, and their characteristics.}
    \label{tab:exp_cohorts}
\end{table*}

An A/B test was conducted to evaluate the impact of nudging over 12 weeks versus no nudging on subsequent steps and MVPA behavior. A total of 169,667 users were randomly selected from the existing pool of participants in the physical activity program, which comprised 14,901 participants who were only enrolled in the physical activity program (Group 1), and 154,766 who were enrolled in both the physical activity and nutrition programs (Group 2). Because the underlying characteristics of these groups were different, they were analyzed as two separate groups to ensure fair comparison of results. Each group of participants were further split into two cohorts of roughly equal size, and equally matched on key characteristics including demographics, baseline behaviors (before the start of nudging), mobile OS, and brand of fitness trackers. These cohorts were named "Nudges" and "No Nudges". Table \ref{tab:exp_cohorts} summarizes the number of participants and their characteristics in each group and cohort.

Program subject matter experts authored a total of 96 nudges in the Nudge Library, comprising 31 nudges targeting steps as the behavior goal, and 65 targeting MVPA. During the 12-week study period, NudgeRank\texttrademark{} generated 1 nudge per day for participants in the "Nudges" groups only, which were delivered via push notifications to the mobile health application. Participants in the "No Nudges" groups did not receive any nudges.

All participants continued to be exposed to the usual communications and mass marketing efforts conducted by the respective programs. Besides the experimental variable of nudges versus no nudges, there were no other differences in how they were treated.

The daily steps and MVPA behavior of participants were collected during the 12 week study via the mobile health application, to evaluate the impact of nudging on those behaviors. One-tailed independent samples $t$-tests were performed to evaluate the differences between the "Nudges" and "No Nudges" groups.

The study design was approved by the relevant authority in Singapore, and all participants had previously given consent for their fitness data to be collected, for their data to be used to send marketing collateral and personalized messages, nudges or notifications, and for evaluation of results.

\subsection{Results}

During the 12-week period, a total of 1.12 million nudges were sent, achieving an open rate of 13.1\% (147k nudges), which was significantly higher than existing baseline open rates of about 4\%. Out of the nudges opened, 6 times more nudges were rated useful (11.7\%) versus not useful (1.9\%), suggesting that the nudges were very well received by participants.

The behavior data collected revealed a significant impact of digital nudging in increasing physical activity. Group 1 "Nudges" participants performed  6.17\% more daily steps (3,088 vs. 2,909 steps; $p = 3.09\times10^{-4}$ and 7.61\% more weekly minutes of MVPA (50.6 vs. 47.0 minutes; $p = 1.16\times10^{-2})$ than Group 1 "No Nudges" participants. The results were statistically significant at $\alpha = 0.05$, indicating that the nudges were effective in increasing both steps and MVPA behavior.

The effects on Group 2 participants, though smaller, were still statistically significant at $\alpha = 0.05$. Group 2 "Nudges" participants performed 1.89\% more daily steps (3,900 vs. 3,827 steps; $p = 2.85\times10^{-4}$) and 2.46\% more weekly minutes of MVPA (88.4 vs. 86.2 minutes; $p = 2.02\times10^{-2})$ than Group 2 "No Nudges" participants. We noted that Group 2 participants were already naturally more physically active than Group 1 participants (refer to their mean daily steps and MVPA at start of experiment, in Table \ref{tab:exp_cohorts}), so perhaps the opportunity to drive behavior change of a similar magnitude was smaller.

\begin{figure}
    \centering
    \includegraphics[width=1\linewidth]{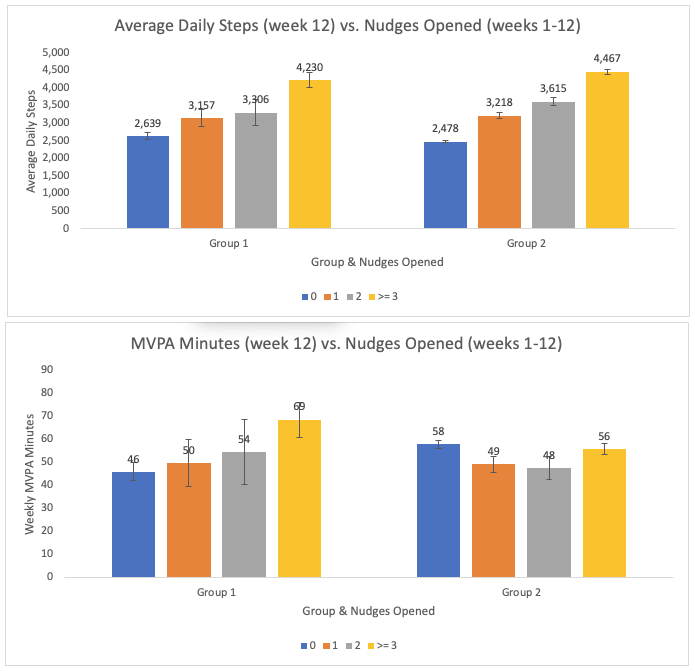}
    \caption{Dose response relationship between number of nudges opened and physical activity (steps and MVPA).}
    \Description{Bar charts showing that the more nudges opened, the more steps and MVPA performed by participants in the A/B test.}
    \label{fig:dose-response}
\end{figure}

The results also revealed a dose response relationship between the number of nudges opened by "Nudges" participants during the 12 weeks and physical activity (Figure \ref{fig:dose-response}). Participants who opened more nudges tended to walk more steps during the 12 weeks. For MVPA, this relationship was more apparent in Group 1 than in Group 2.

Overall, the results of this study provided strong evidence that algorithmic digital nudging can effect behavior change (physical activity in this case) across a large and diverse population. Nudging can be further extended to other types of health behaviors like nutrition, sleep and smoking cessation, with the potential to drive tremendous benefits in population health outcomes.

\end{document}